\documentclass[sn-standardnature,iicol]{sn-jnl}

\usepackage[utf8]{inputenc}
\usepackage{graphicx}
\usepackage{siunitx}
\usepackage{xcolor}
\usepackage{placeins}
\usepackage{textgreek}

\jyear{2022}%

\begin{document}

\title[ ]{Carrier-envelope phase controlled dynamics of relativistic electron beams in a laser-wakefield accelerator}

\author*[1]{\fnm{Lucas} \sur{Rovige}}\email{lucas.rovige@ensta-paris.fr}
\author[1]{\fnm{Joséphine} \sur{Monzac}}
\author[1]{\fnm{Julius} \sur{Huijts}}
\author[1]{\fnm{Igor A.} \sur{Andriyash}}
\author[1]{\fnm{Aline} \sur{Vernier}}
\author[1]{\fnm{Jaismeen} \sur{Kaur}}
\author[1]{\fnm{Marie} \sur{Ouillé}}
\author[1]{\fnm{Zhao} \sur{Cheng}}
\author[2]{\fnm{Vidmantas} \sur{Tomkus}}
\author[2,3]{\fnm{Valdas} \sur{Girdauskas}}
\author[2]{\fnm{Gediminas} \sur{Raciukaitis}}
\author[2]{\fnm{Juozas} \sur{Dudutis}}
\author[2]{\fnm{Valdemar} \sur{Stankevic}}
\author[2]{\fnm{Paulius} \sur{Gecys}}
\author[1]{\fnm{Rodrigo} \sur{Lopez-Martens}}
\author[1]{\fnm{Jérôme} \sur{Faure}}

\affil*[1]{\orgdiv{LOA}, \orgname{CNRS, Ecole Polytechnique, ENSTA Paris, Institut Polytechnique de Paris}, \orgaddress{\street{181 Chemin de la Hunière}, \city{Palaiseau}, \postcode{91120}, \country{France}}}
\affil[2]{\orgname{Center for Physical Sciences and Technology}, \orgaddress{\street{Savanoriu Ave. 231}, \city{Vilnius}, \postcode{LT-02300}, \country{Lithuania}}}
\affil[3]{\orgname{Vytautas Magnus University}, \orgaddress{\street{K.Donelaicio St. 58}, \city{Kaunas}, \postcode{LT-44248}, \country{Lithuania}}}

\abstract{In laser-wakefield acceleration, an ultra-intense laser pulse is focused into an underdense plasma in order to accelerate electrons to relativistic velocities. In most cases, the pulses consist of multiple optical cycles and the interaction is well described in the framework of the ponderomotive force where only the envelope of the laser has to be considered. But when using single-cycle pulses, the ponderomotive approximation breaks down, and the actual waveform of the laser has to be taken into account. In this paper, we use near-single cycle laser pulses to drive a laser-wakefield accelerator. We observe variations of the electron beam pointing on the order of 10\,mrad in the polarisation direction, as well as 30\% variations of the beam charge, locked to the value of the controlled laser carrier-envelope phase, in both nitrogen and  helium plasma. Those findings are explained through particle-in-cell simulations indicating that low-emittance, ultra-short electron bunches are periodically injected off-axis by the transversally oscillating bubble associated with the slipping carrier-envelope phase.}


\maketitle

\section{Introduction}
Laser-wakefield acceleration \cite{tajima_laser_1979} is a promising way to accelerate electrons to high energy over a very short distance due to the extreme accelerating fields that can be sustained by non-linear plasma waves. Such laser-plasma accelerators have demonstrated acceleration of multi-GeV beams on a few tens of centimetres length \cite{gonsalves19}. These high energy electron beams are good candidates for providing high-brilliance ultrashort secondary X-ray sources via betatron radiation \cite{rousse_production_2004}, Compton scattering \cite{TaPhuoc2012,Chen2013}, or free-electron lasers \cite{Wang2021}.  In most cases, the laser pulses used to drive the LWFA are sufficiently long so that the interaction with the plasma can be described through the widely used cycle-averaged ponderomotive framework \cite{mora_kinetic_1997}, where the physics depends only on the envelope of the pulse, and is polarization independent.  But in high-repetition rate laser plasma accelerators capable of producing kilohertz electron beams at MeV energies \cite{guenot_relativistic_2017,rovige_demonstration_2020,salehi2021laser}, shorter, sub-5\,fs pulses are used. These high-repetition rate MeV electron beams could be used in ultrafast electron diffraction experiments \cite{he_electron_2012,he_capturing_2016,faure_concept_2016}, radiation hardness assessment \cite{hidding_laser-plasma_2017}, radio-biology \cite{Rigaud2010,lundh2012,cavallone2021} or for ultrashort positron beam generation \cite{audet2021}. 
When a near-single cycle pulse drives a wakefield, the ponderomotive approximation breaks down and asymmetries arise in the electron dynamics. In this case, the actual waveform taking into account the phase between the envelope and the carrier wave (the CEP) of the laser pulse needs to be considered. Nerush et al. \cite{nerush_carrier-envelope_2009} showed that these asymmetries are caused by higher-order terms of the plasma response and are CEP dependent. Moreover, as the laser pulse propagates in the plasma, the carrier-envelope phase of the pulse slips due to plasma dispersion on a typical length scale of \cite{nerush_carrier-envelope_2009,faure_review_2018} :
\begin{equation}
L_{2\pi} = \frac{c}{v_{\phi}-v_g}\lambda_0 
\end{equation}
where $v_{\phi}$ is the laser phase velocity, $v_g$ the laser group velocity, c the light velocity, and $\lambda_0$ the laser central wavelength.
This slippage of the CEP causes the wake (or bubble) to oscillate transversally during the propagation on the length scale $L_{2\pi}$, which has been shown to trigger periodic off-axis injection of electron bunches in simulations \cite{huijts2021identifying,xu_periodic_2020,kim21_cep} and in our recent experiment \cite{huijts22}. Our previous work demonstrated clear variations of the experimental beam pointing in the polarization direction with the CEP, attributed to the bubble transverse oscillation injection. But the impact of this mechanism on other beam parameters such as charge or energy spectrum still required further study. Additionally, the experiment was carried out in a nitrogen plasma where CEP-dependent ionization injection \cite{lifschitz_optical_2012} can occur, even though our simulations indicated this effect remained marginal. Using a low-Z gas such as helium would strictly rule out ionization injection and therefore confirm our interpretation. \par
In this paper, we observe CEP-dependent variations in both beam pointing and charge, first in nitrogen with an enhanced level of control through polarization, and then in  helium. The injection and subsequent dynamics of the electron bunches, as well as couplings with the laser spot asymmetry are explored in further details through particle-in-cell simulations. We describe the experimental setup in Sec.\,\ref{sec:exp_set}, the experimental results are presented in in Sec.\,\ref{sec:exp_res} and PIC simulations are discussed in Sec.\,\ref{sec:simu}.

\section{Experimental setup}
\label{sec:exp_set}

\begin{figure}[t!]
    \includegraphics[width=0.45\textwidth]{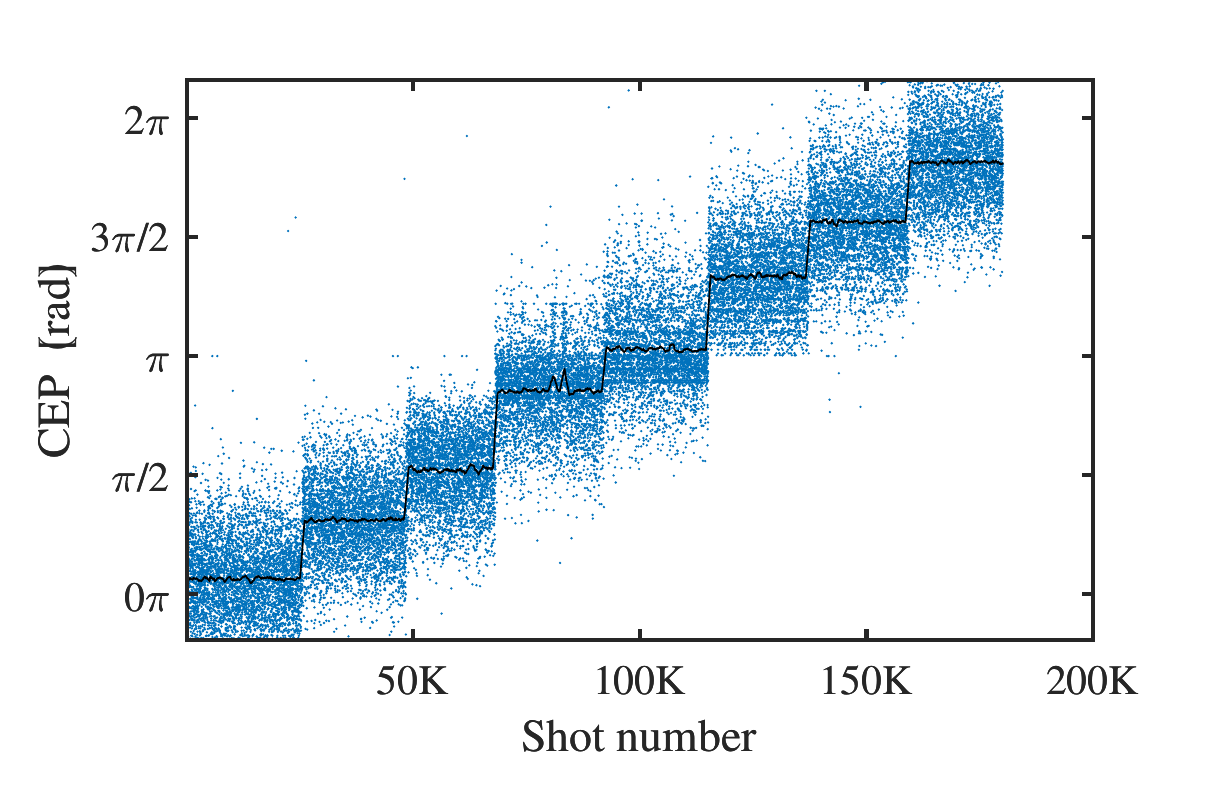}
    \caption{\label{fig:CEP} Single shot CEP measurement (blue) acquired by the Fringeezz by varying the target value between 0 and $\mathrm{2\pi}$, and moving average value on 100 shots (black).}
\end{figure}

The laser-wakefield accelerator is driven by a kilohertz laser delivering 4.0\,fs (1.5 optical cycle at 800\,nm), 3.0\,mJ pulses \cite{faure_review_2018,guenot_relativistic_2017-1,rovige_demonstration_2020} focused by an f/2 off-axis parabola to a 5\,$\mathrm{\mu m}$ spot, reaching a peak intensity in vacuum I = $\SI{1.9e18}{\watt\per\square\centi\meter}$. The near-single cycle pulses are obtained by postcompressing 10\,mJ, 23\,fs pulses by combining spectral broadening in a Helium-filled hollow-core fiber with compression by chirped mirrors.\cite{bohle2014,ouille_relativistic-intensity_2020}. The laser is focused $\SI{10}{\micro\meter}$ before the center of a supersonic continuously flowing $\mathrm{N_2}$ or He gas jet, at a distance of $\SI{150}{\micro\meter}$ from the exit of the ``de Laval" nozzle \cite{schmid2012supersonic} with a $\SI{60}{\micro\meter}$ throat and $\SI{180}{\micro\meter}$ exit diameter.  We use a differential pumping system where the nozzle flows into a centimetre-size chamber, with millimetre-size entrance holes for the laser, pumped by a separated primary pump while the large experimental chamber is pumped by a root and a turbo-molecular pump. This system keeps the residual pressure below \num{e-2}\,mbar in the large chamber even with continuously flowing Helium at high pressures. The gas density is measured by illuminating the gas jet from the side with a white LED lamp, and imaging it on a quadri-wave latereral shearing interferometer (QWLSI) \cite{primot1995achromatic,primot2000extended} which measures the phase shift induced by the gas. Molecular gas density is then retrieved from the 2D phase map via Abel inversion, assuming radial symmetry along the nozzle axis. The plasma density is then deduced assuming ionization of nitrogen up to $\mathrm{N^{5+}}$, a molecule of $\mathrm{N_2}$ yielding therefore 10 electrons, or total ionization of He into $\mathrm{He^{2+}}$, releasing 2 electrons in this case. In section \ref{sec:n2} pure nitrogen with a backing pressure of 13 bar is used, which corresponds to a peak plasma density $\mathrm{n_e} = \SI{8.3e19}{\per\cubic\centi\meter}$. In section \ref{sec:he}, pure helium with backing pressure of 85 bar and 110 bar is used, corresponding to plasma densities of respectively $\mathrm{n_e} = \SI{1.0e20}{\per\cubic\centi\meter}$ and $\mathrm{n_e} = \SI{1.3e20}{\per\cubic\centi\meter}$. A probe pulse, obtained from the main beam through a holed mirror, is used to image the plasma from the side. \par
In section \ref{sec:n2}, a zero-order broadband half-wave plate is inserted in the laser beam in order to control the linear polarization direction, but due to the extremely large laser spectrum ranging from 500\,nm to 1000\,nm, the waveplate was found to add some residual ellipticity, of the order of 1\% for the vertical direction (incident polarization very close to the plate axis) and 10\% for the horizontal one. \par
The electron beam charge, spatial profile and pointing is measured with a 3" Ce:YAG scintillator and imaged on a CCD camera. The electron energy spectrum is measured by inserting a motorized spectrometer in the beam, composed of a set of two 25\,mm diameter circular magnets providing a peak 117\,mT magnetic field and a horizontal line of $\SI{500}{\micro\meter}$ pinholes.\par

The laser CEP is stabilized and controllable, with a typical stability of 500\,mrad RMS. The electron data presented in this paper are averaged over 100 shots, which brings down the typical CEP variations to 50\,mrad for these integrated measurements (see Fig.\,\ref{fig:CEP}). The CEP stabilization is done in two loops: the first one controls the CEP of the oscillator, by modulating the laser pump power (performed with a XPS800 by Menlo Systems), while the second feedback loop is done after the amplification and post-compression stages, where a wedge reflection is sent to a f-2f interferometer \cite{kakehata2001single}. Each single-shot interference pattern is measured and analysed by a Fringeezz \cite{lucking2014approaching} (Fastlite) which is then fed back to an acousto-optic modulator. The Fringeezz allows only for relative CEP measurement, meaning the values given here are linked to the actual laser CEP by an unknown arbitrary constant.  \par
Each data point in beam pointing, charge and spectrum corresponds to the average over 20 measurements, each of which is obtained by accumulating 100 consecutive shots. The errorbars correspond to the standard deviation on these 20 measurements. The beam profiles are obtained by acquiring over 100 consecutive shots.\par

\begin{figure*}
    \includegraphics[width=0.9\textwidth]{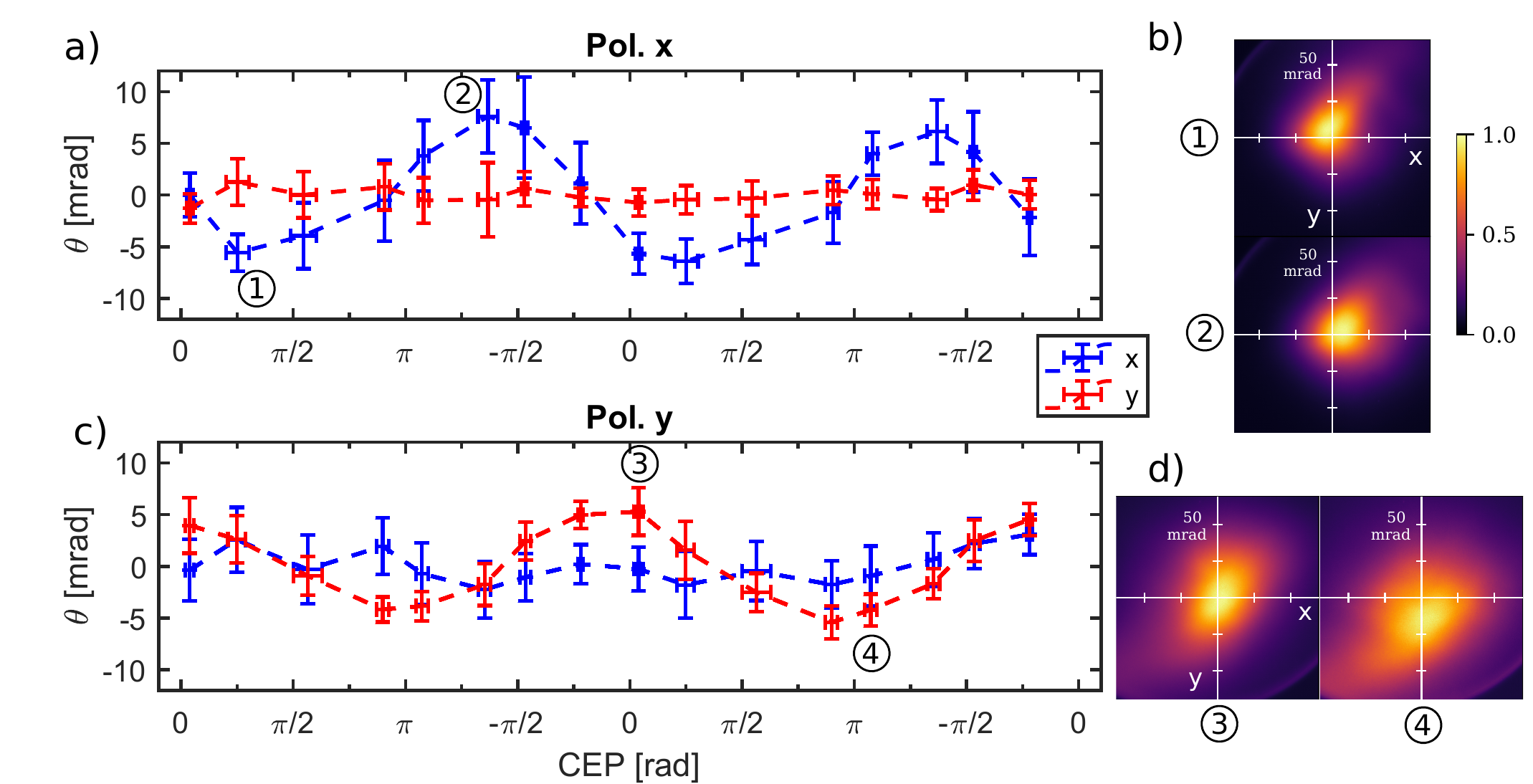}
    \caption{\label{fig:HV} \textbf{CEP effects on the electron beam depending on the laser polarization with $\mathrm{N_2}$ gas.} a) Electron beam pointing in both x (blue) and y (red) directions for a horizontal laser polarization (along x). Each point corresponds to 2000 shots. b) Images of the beams accumulated on 100 shots corresponding to CEPs of $\mathrm{\pi}/4$  and $\mathrm{-3\pi}/4$ in the horizontal case. c) Electron beam pointing in both x and y directions for a vertical laser polarization (along y).  d) Images of the beams corresponding to CEPs of 0 and $\mathrm{\pi}$ in the vertical case.  }
\end{figure*}

\section{Experimental results}
\label{sec:exp_res}
\subsection{Polarization control in nitrogen}
\label{sec:n2}
Here, we operate in similar conditions as in \cite{huijts22}, by using pure nitrogen gas with a plasma density $\mathrm{n_e} = \SI{8.3e19}{\per\cubic\centi\meter}$, with the added feature that we can now control the laser linear polarization direction by rotating a half-wave plate placed in the beam. A clear periodic dependence of the electron beam pointing on the CEP is observed, as shown in Fig.\,\ref{fig:HV}. The beam oscillates in the laser polarization direction, with an amplitude of around 5 mrad, while it does not move in the perpendicular direction. When the polarization of the laser is rotated by $90^\circ$, so is the direction of oscillations of the beam pointing, indicating we can achieve fine control of the beam pointing in all the directions. Additionally, Fig. \ref{fig:HV}b,d show a significant difference in divergence between the two cases (43 mrad vs 68 mrad), which can be explained by the fact that the beam was optimized by adjusting experimental parameters such as chirp and position of the focus for the horizontal polarization direction, which were then kept constant when shifting to vertical polarization while a residual dispersion could be induced by the rotated waveplate. \par

\begin{figure}
    \includegraphics[width=0.48\textwidth]{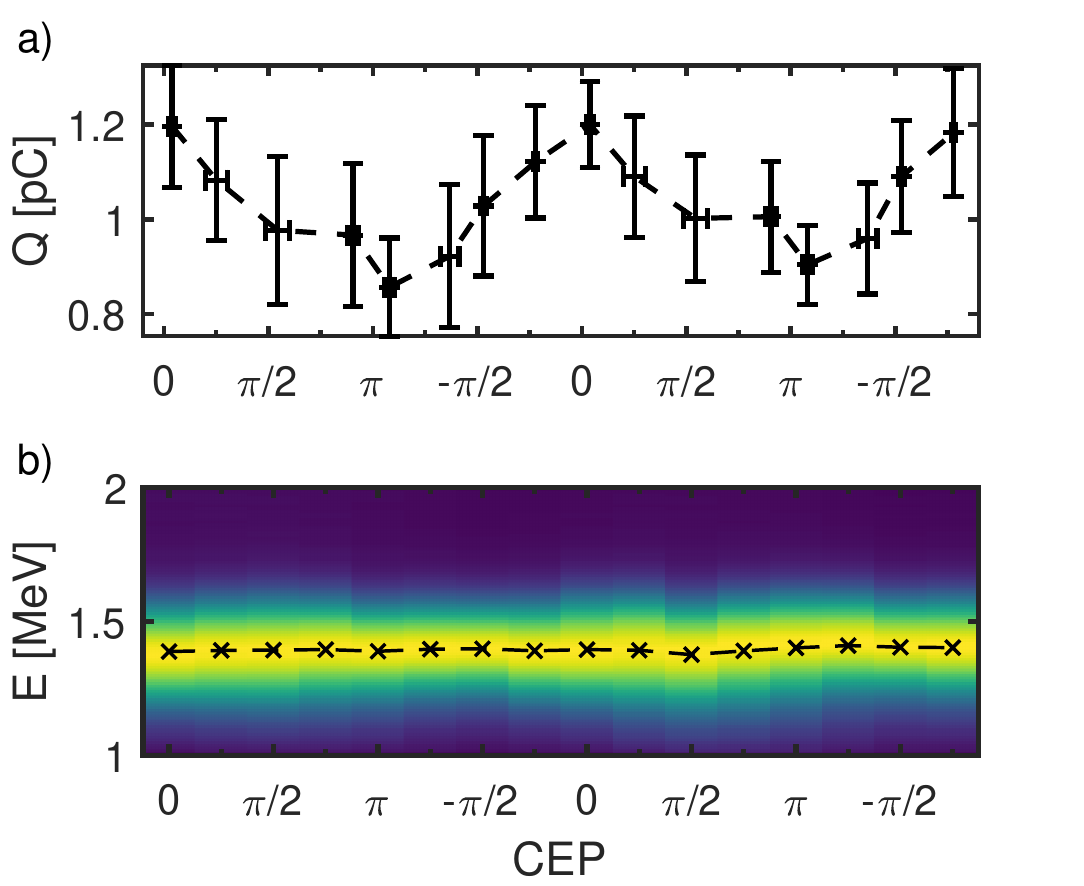}
    \caption{\label{fig:HQE} a) Electron beam charge Q and b) electron spectra plotted against CEP for the horizontal polarization case in $\mathrm{N_2}$. The black crosses correspond to the mean energy. Each points corresponds to 2000 shots.}
\end{figure}

In our previous experiment \cite{huijts22}, a slight beam charge variation depending on the CEP could be inferred, but it was dominated by higher amplitude CEP-independent charge variations during the scan. Moreover, the measured electron energy spectrum oscillated moderately with CEP, but the spectrometer geometry where the moving beam was sampled by a fixed pinhole did not allow to discriminate between an actual effect of the CEP on the spectrum, or just a difference of energy depending on the position sampled in the electron beam. Figure \ref{fig:HQE}, showing the charge and energy dependence on CEP with the horizontal laser polarization gives us new insight on these issues. Indeed, Fig.\,\ref{fig:HQE}a demonstrates periodic dependence of the charge on the CEP, with a significant total variation of around 30\%. Additionally, the horizontal laser polarization enable the use of a series of pinholes on a horizontal line, which permits to obtain information on the electron spectra independently from the CEP-dependent beam pointing by averaging the spectra on the whole sampling line. With this new geometry, Fig.\,\ref{fig:HQE}b shows no clear effect, at least in this case, of the laser CEP on the energy spectrum. 
   
\subsection{Carrier-envelope phase effects in helium}
\label{sec:he}
\begin{figure*}
    \includegraphics[width=0.95\linewidth]{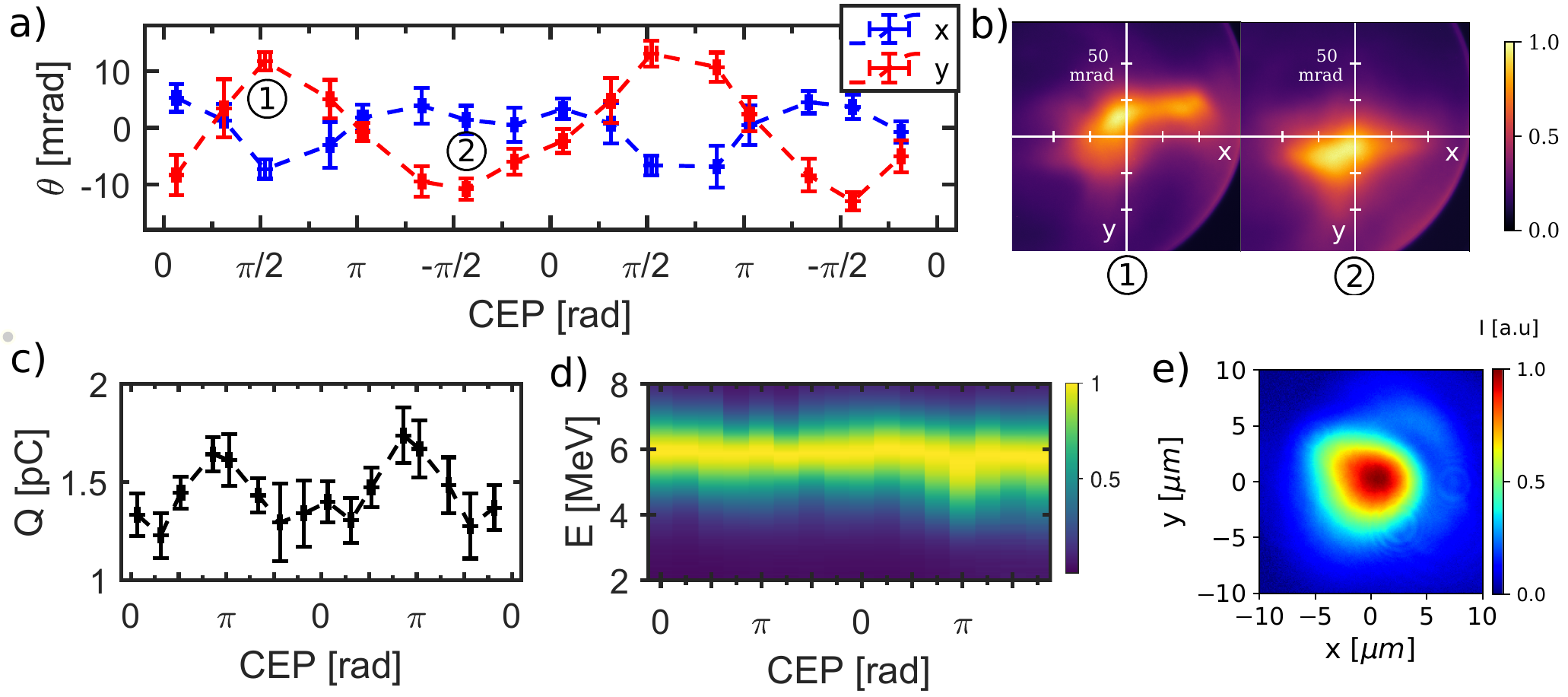}
    \caption{\label{fig:he85} \textbf{CEP effects on the electron beam in helium with $\mathbf{P_{back}=}$\,85 bar, corresponding to a plasma density \boldmath${\mathrm{n_e} = 1.0 \times 10^{20}\,\mathrm{cm^{-3}}} $} a) Electron beam pointing in both x (blue) and y (red) directions (laser polarization along y) plotted against CEP. b) Images of the beams corresponding to CEPs of $\mathrm{\pi}/2$  and $\mathrm{-\pi}/2$.  c) Electron beam charge Q against the laser CEP  d) Electron spectra against CEP.  e) Laser focal spot in vacuum.}
\end{figure*}
In our previous work \cite{huijts22}, experiments were carried out solely in nitrogen to ensure high electron density while limiting the pumping load. While nitrogen is a high-Z gas that theoretically allows for ionization injection to occurs, our simulations indicated that the laser intensity remained too low throughout the interaction for this process to be significant. The beam pointing variations were therefore attributed to the CEP-dependent oscillations of the plasma wave asymmetry \cite{nerush_carrier-envelope_2009} which can trigger off-axis injection of electrons in the plasma bubble \cite{huijts2021identifying,kim21_cep,huijts22}. Carrying out the experiment in pure helium would allow to completely isolate this process unrelated to ionization injection, as helium is completely ionized at an intensity $I\simeq \SI{e16}{\watt\per\centi\meter}$ which is reached very early in the front of the pulse. Moreover, using a low-Z gas yields better performances for the LPA. Indeed, laser propagation leads to much lower ionization defocusing in helium than in nitrogen, allowing to reach higher intensities and longer propagation distances and therefore higher electron energies. The use of the differential pumping system described in Sec.\,\ref{sec:exp_set} enabled the use of continuously flowing helium with the high pressure necessary for this study. The half-wave plate was removed because it was causing a performance loss of the accelerator, most probably by adding spectral phase imperfections due to the  broad laser spectrum. The polarization of the laser is therefore vertical (along y) in this section.\par
As anticipated, using helium as a gas target for our LPA allows us to accelerate electrons to significantly higher energies, with spectra peaked around 6\,MeV (see Fig.\,\ref{fig:he85}d). The beam parameters exhibit a CEP-dependence that is similar to what was obtained in nitrogen, with the beam pointing oscillating in the polarization direction (y) with a large amplitude of 11\,mrad, as shown in Fig.\,\ref{fig:he85}. A surprising observation is that, the beam pointing in the perpendicular direction also varies periodically with the CEP, but with a smaller amplitude ($\sim$5\,mrad). This is surprising because theoretically the plasma asymmetry, and therefore the CEP-dependence, should only occur in the plane of polarization \cite{nerush_carrier-envelope_2009}. We explain this unexpected behaviour by a coupling between the asymmetry in the polarization plane, and an asymmetry in the perpendicular plane due to an asymmetric focal spot, as can be seen in Fig.\,\ref{fig:he85}e. This is studied in more details with PIC simulations in section \ref{sec:perpasym}. Additionally, the beam charge varies by up to 30\%, between 1.2\,pC and 1.7\,pC, due to CEP changes, while the energy spectrum remains unaffected (see Fig.\,\ref{fig:he85}c,d). These results confirm the explanation of the observed CEP effects being due to the off-axis injection of electrons caused by the bubble asymmetry, as ionization injection cannot occur at all in helium. \par

\begin{figure}
    \includegraphics[width=0.48\textwidth]{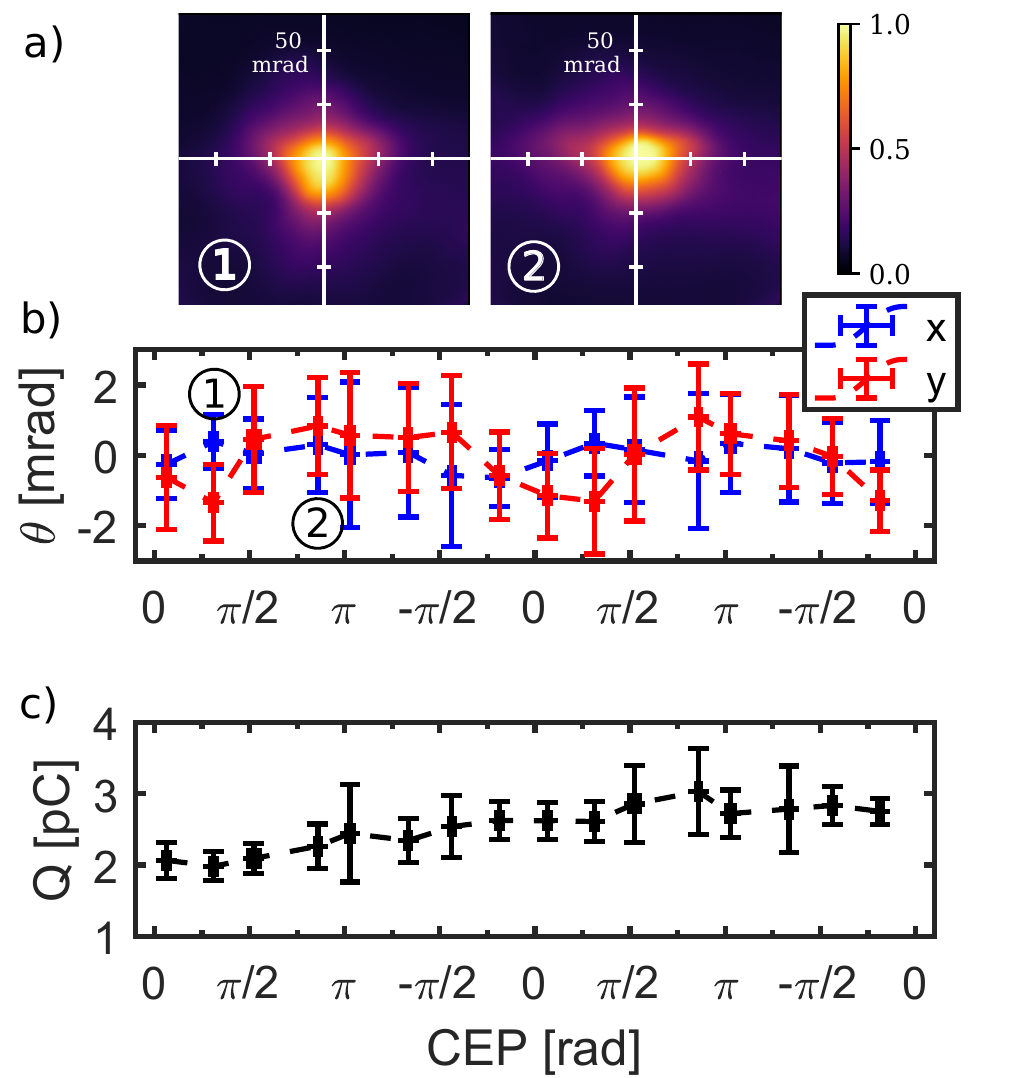}
    \caption{\label{fig:he110} a) Beam profiles for two different CEP of $\pi/4$ and $3\pi/4$. Typical divergence is 40\,mrad FWHM. b) Electron beam pointing $\theta$ and c) electron beam charge plotted against CEP, in helium with $\mathrm{P_{back}} = 110$ bar, corresponding to a plasma density $\mathrm{n_e} = \SI{1.3e20}{\per\cubic\centi\meter}$.}
\end{figure} 
When increasing the plasma density to $\mathrm{n_e} = \SI{1.3e20}{\per\cubic\centi\meter}$, CEP effects on the electron beam are reduced, resulting in much smaller pointing variations of 1\,mrad amplitude, while charge appears this time uncorrelated to CEP changes (Fig.\,\ref{fig:he110}) This higher density case yields more charge per shot than the previous one, with charge slowly evolving from 2\,pC/shot  to 3\,pC/shot during the scan, seemingly uncorrelated to CEP. The energy of the electrons is slightly lower, with a spectra peaked at 5\,MeV.    

These new experimental results are interesting from both a technical and fundamental point of view. The experiment in nitrogen reproduces quite well the results previously obtained in \cite{huijts22} while adding another layer of control through the polarization : the beam direction can now be controlled through a polar and azimuthal angle ($\theta,\phi$), where $\theta$ is controlled by changing the CEP value and $\phi$ by rotating the waveplate. Moreover, we have been able to observe similar CEP effects using a helium gas jet, confirming our explanation and excluding ionization injection. 

\section{PIC simulations and discussion}
\label{sec:simu}

\subsection{Particle-in-cell simulations in helium with an ideal laser driver}
\label{sec:simsym}

\begin{figure*}
    \includegraphics[width=0.95\linewidth]{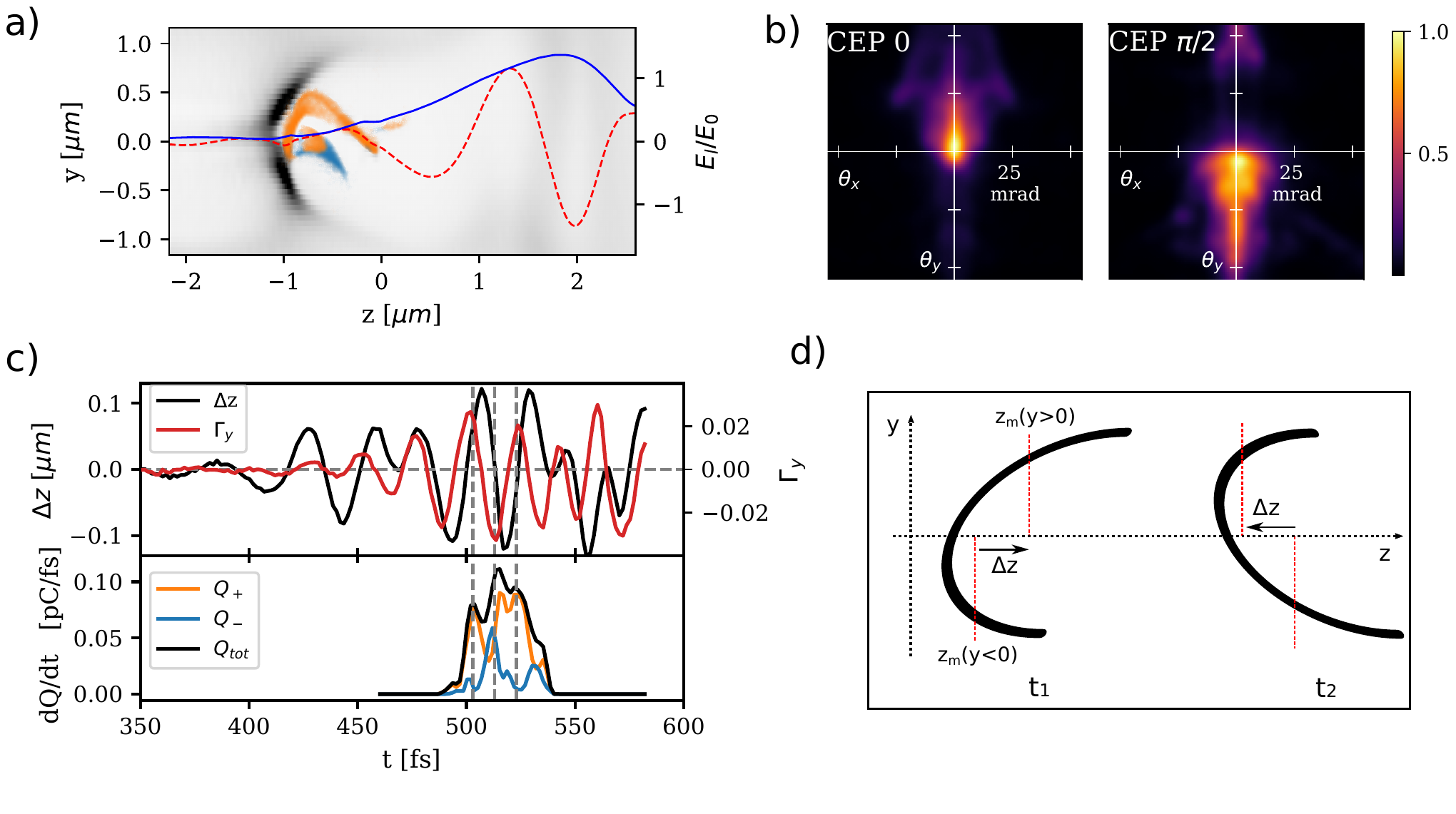}
    \caption{\label{fig:sim_sym} \textbf{Particle-in-cell simulation of a LPA driven by a 4.0\,fs laser in a helium plasma.}  a) Snapshot of the wakefield for a initial CEP of 0, showing three different injected sub-bunches. Electron density in the z-y plane is shown in gray, and injected electrons are displayed in orange (blue) when their pointing is positive (negative) at the end of the simulation. The normalized laser electric field $E_l/E_0=E_l/(m_ec\omega_0/e)$ (red dashed line) and its envelope (blue solid line) are also shown.  b) Output beams from simulations with initial CEP of 0 and $\pi/2$.  c) Asymmetry of the wakefield in the y-direction (red) and difference in the mean position of the bubble in z between the top part (y$>$0) and the bottom part (y$<$0) of the wake ($\Delta z$, black). Charge injection rate for the two electron populations shown in a) with corresponding colors, as a function of the simulation time for an initial CEP of 0. The gray dashed lines highlight the three main injection events.   d) Schematic description of the longitudinal bubble asymmetry for two times $t_1$ and $t_2$ corresponding to $\Delta z <0$ and $\Delta z >0$.  The change of asymmetry from $t_1$ to $t_2$ would lead to injection of electrons from the top of the bubble.}
    
\end{figure*}

For a more detailed understanding of the interaction between the near-single cycle pulse and the plasma leading to the observed CEP effects, we performed particle-in-cell simulations with the spectral quasi-cylindrical code FBPIC \cite{lehe_spectral_2016}. In the simulations the mesh is $\Delta z = \lambda_0/48$, $\Delta r = 3\Delta z$, and five azimuthal Fourier modes are used to properly describe the asymmetric physics. The simulation is initialized with neutral helium and ionization is computed with the Ammosov-Delone-Krainov model of tunnel ionization \cite{ammosov_tunnel_1986}. Helium is initialized with 256 neutral macro-particles per r-z cell, and each helium atom can produce 2 electrons after complete ionization. The plasma profile is a super-gaussian matching experimental measurements, with a peak plasma density $n_e = \SI{1.35e20}{\per\cubic\centi\meter}$. In this section, we use an idealised gaussian temporal and spatial profile (with respectively a FWHM and waist matching the experiment) to model our laser driver and identify the mechanisms leading to carrier-envelope phase effects. The laser focus is positioned $\SI{10}{\micro\meter}$ before the peak density, and a positive chirp of $\SI{10}{\square\femto\second}$ is used to pre-compensate the dispersion due to the propagation in the plasma. The energy per pulse is 2.8\,mJ, which was slightly adjusted to better match experimental results. The laser is linearly polarized in the vertical (y) direction.\par

The accelerated electron beam in our simulations has a charge of 3\,pC, and an energy spectrum peaked around 7\,MeV, matching quite well the experimental results. The simulations show a similar behaviour as in \cite{huijts22}, where evolving CEP-dependent asymmetry of the plasma wave is responsible for off-axis injection of several electrons sub-bunches (see Fig.\,\ref{fig:sim_sym}a). To study it in more details, we characterize this asymmetry with two quantities, the transverse centroid of the bubble normalized to the laser waist : $\Gamma_y = \frac{\int n_e y dy}{\int n_e dy}\times\frac{1}{w_0}$ and the difference in the mean longitudinal position of the bubble between the $y>0$ and $y<0$ parts: $\Delta z = z_{m}(y>0)-z_{m}(y<0)$, with $z_m = \frac{\int n_e z dz}{\int n_e dz}$ considering only densities higher than $0.1n_c$. This longitudinal asymmetry is illustrated by Fig.\,\ref{fig:sim_sym}d.  As shown in Fig.\,\ref{fig:sim_sym}c, the asymmetry of the wake oscillates during the propagation in the plasma due to the shifting CEP. It appears that $\Delta z$  is dephased by $\pi/2$ with respect to $\Gamma_y$, meaning that the bubble is not only moving upwards and downwards as described in the analytical work of Kim \textit{et al.} \cite{kim21_cep}, but it also rotates around its center, i.e. it is moving forward and backward alternately for positive and negative y, as schematically represented in Fig.\ref{fig:sim_sym}d. By comparing the charge injection rate to the quantities characterising the asymmetry in Fig.\,\ref{fig:sim_sym}c, it appears that electrons are injected when the bubble is moving backward on one side, with the injection peaks occurring when $\Delta z$ changes sign, corresponding to its maximum local slowdown of one side of the bubble, but also to an extremum of the transverse asymmetry. This is analogous to the injection in a density gradient where the bubble moves backwards and traps electrons \cite{bulanov_particle_1998,tomassini_2003}. Because the bubble is moving backwards only on one side at a time, electrons are injected off-axis on an alternating y-side depending on the CEP. When the electrons are injected from the bottom, they have a non-zero positive transverse momentum and therefore end up pointing mainly upward after a single betatron oscillation at the end of the simulation. Conversely, electrons end up pointing downward when injected from the top. In the direction perpendicular to the polarization, the plasma wave remains perfectly symmetric, electrons are injected on-axis and therefore the beam is centred in this direction.
Moreover, during its interaction with the plasma, the laser is strongly redshifted up to a wavelength of $\SI{1.4}{\micro\meter}$, which reduces the CEP oscillation period, increases the amplitude of the asymmetry through the scaling $\Gamma_y \propto a_0^3\lambda^4_0$ \cite{nerush_carrier-envelope_2009} and facilitates electron trapping by lowering the wake phase velocity \cite{faure_review_2018,beaurepaire_electron_2014}.\par
The sub-bunches have very interesting properties: their injection is very localised, which translates into an extremely short bunch duration between 1-2\,fs. Additionally, the trapping mechanism causes them to have a narrow transverse momentum spread, and then to perform collective betatron oscillation in the bubble. Due to their localised injection and low spread in $P_y$, the emittance of the sub-bunches is relatively small, of the order of $\SI{100}{\nano\meter}$.

\begin{figure}
    \includegraphics[width=0.48\textwidth]{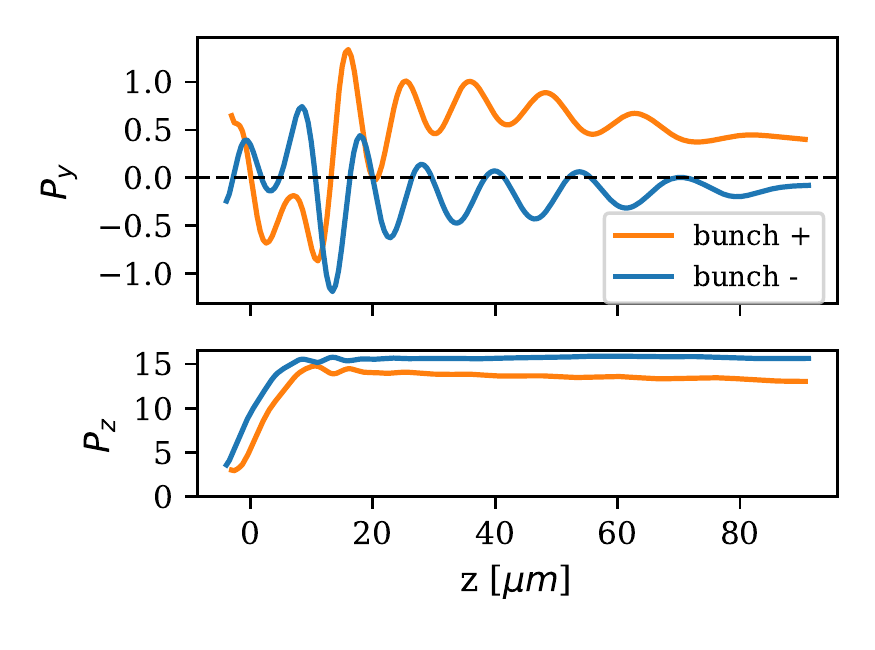}
    \caption{\label{fig:track} Top : transverse momentum in the polarization direction as a function of propagation distance for two sub-bunches injected with a positive (bunch\,+) and negative (bunch\,-) initial transverse momentum, corresponding to opposite signs of $\Gamma_y$ at the moment of injection. Bottom: longitudinal momentum of the two sub-bunches. Both momenta are normalized to $\mathrm{m_ec}$.}
\end{figure} 
Figure\,\ref{fig:sim_sym}b also shows that, for an initial CEP of 0, significantly more injected electrons end up pointing upwards than downwards, resulting in a positive total beam pointing, because, as can be seen in Fig.\,\ref{fig:sim_sym}c, a part of the electrons from the second injection, which occurs from the top and should point downward, actually ends up with positive pointing. To explain this, we track in Fig.\,\ref{fig:track} two sub-bunches, injected at opposed transverse asymmetry values, during their propagation. We will call `bunch +' an electron sub-bunch injected with a positive transverse momentum in the polarization direction ($P_y>0$) from the bottom of the bubble ($y<0$) and `bunch -' an electron sub-bunch injected with a negative transverse momentum in the polarization direction ($P_y<0$) from the top of the bubble ($y>0$). The transverse and longitudinal momenta of the two bunches are plotted in Fig.\,\ref{fig:track}. The two bunches experience a similar trajectory, where they are injected with a non-zero transverse momentum, perform a single betatron oscillation before catching up with the laser and then oscillate in the laser electric field. Longitudinally, the electrons are accelerated over a distance of $\SI{13}{\micro\meter}$ and then keep an approximately stable longitudinal momentum for the rest of their propagation. Apart from the sign of their initial transverse momentum, related to their side of injection, what differentiates the two trajectories is the absolute value of $P_y$ at injection, which is significantly higher in the case of the `bunch\,+' ($P_y$ = 0.58) than the `bunch\,-' ($P_y$ = -0.25), leading to a similar difference in transverse momentum at the end of the simulation. This means that the `bunch\,+' is pointing strongly upwards, while the `bunch\,-' is almost centred, leading to a global positive beam pointing, and explaining why a part of the second injection ends up with a positive (orange) pointing. 
This difference in initial transverse momentum can be caused by the asymmetry of beam-loading effects, because at the moment of injection of the `bunch\,-', the previously injected electron bunch is now on the same side of the bubble after travelling upwards, and perturbs the plasma wave, leading to a reduced $P_y$ at injection. A similar analysis for an initial CEP of $\pi/2$ can be found in Appendix \ref{app:sim}, and leads to global negative beam pointing in the y-direction (see Fig.\,\ref{fig:sim_sym}b).\par

\begin{figure}
    \includegraphics[width=0.48\textwidth]{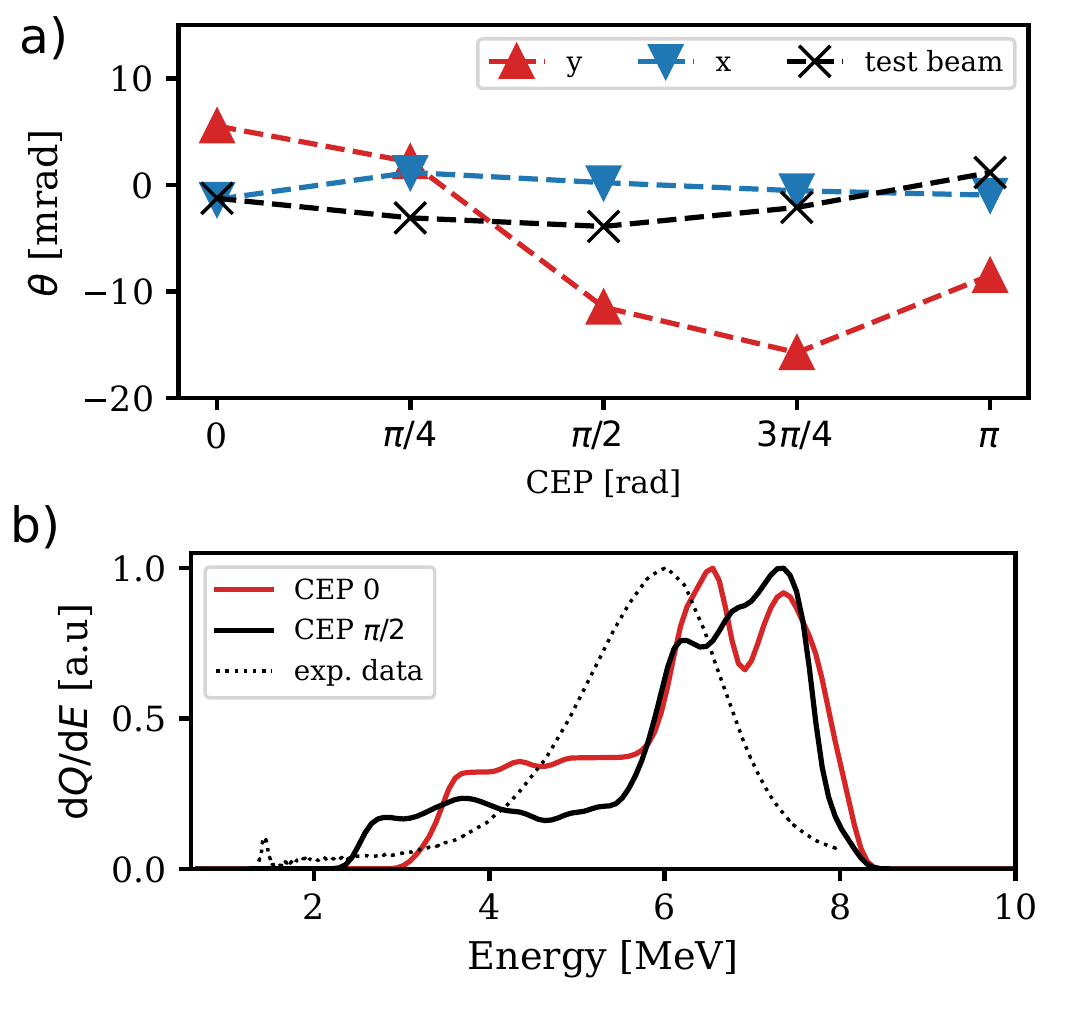}
    \caption{\label{fig:testbeam} a) Beam pointing in the simulations for different CEPs, in the y (polarization) direction (red triangles), in the x direction (blue triangles) and for the test-beam in the y-direction   b) Spectra of the accelerated electrons in the simulations for two different CEPs (solid lines) compared to experimental spectrum (dotted).}
\end{figure} 

We then performed the same PIC simulations by changing the initial CEP between 0 and $\pi$, by increments of $\pi/4$. Because they are the perfect mirror of their $\pi$-shifted positive counterparts, negative CEP values were ignored. In order to validate the role of initial transverse momentum and position of injection of the bunches in their final beam pointing, an electron test-beam was injected on-axis with a zero $P_y$ at the same time as the first real injection occurrence. 
The total charge of the test beam is chosen extremely low (\SI{e-30}{\coulomb}) so as not to interfere with the `physical' electrons. The pointing of the main beam in the simulations shows similar behaviour as observed experimentally, with oscillations in the polarization direction when changing the CEP, and a beam consistently centred in the perpendicular direction (see Fig.\,\ref{fig:testbeam}a).
For the artificially injected test-beam, while a residual CEP-dependence of the beam pointing in the polarization direction to the CEP remains, it is of much lower (4 times) amplitude than for the physical beams, which clearly indicates that the pointing dependence on the CEP observed in the experiments can be attributed to the asymmetric injection, and not to the potentially changing transverse fields during the accelerating process. Still, the residual pointing dependence observed on the test-beam can be explained by the fact that, due to the oscillating asymmetry of the bubble, they are not perfectly on its center when injected and therefore gain some transverse momentum.\par 
The simulations also show only small variations in the electron spectra when changing the CEP, as presented in Fig.\,\ref{fig:testbeam}b, which is in accordance with the experimental results. The energy in the simulations is typically 1\,MeV higher than experimentally measured, which can be explained by the idealised gaussian case used here, which tends to yield better performances, but still remains in good global agreement. The charge variation of 30\% observed experimentally is not reproduced by these simulations, in which only a change in beam charge of the order of 5\% is observed.\par
The experimental results also showed that increasing the plasma density reduces the CEP effects on the measured beam. This can be explained by two factors. Firstly, increasing the plasma density also increases the dispersion in the plasma and therefore results in a smaller CEP slippage length $\mathrm{L_{2\pi}}$. Moreover, the higher density leads to stronger self-focusing of the laser driving a higher amplitude plasma wave in which the time window where injection is possible is larger. So injection is less localised, resulting in an averaging of the effects of the CEP. Additionally, the increased amplitude of the plasma wave facilitates self-injection, making the trapping process less dependent on bubble oscillations (thus on CEP) to occur.\par

These PIC simulations allowed us to reproduce the main experimental feature associated with carrier-envelope phase effects in our experiment, which is the oscillation of the beam pointing in the polarization direction. We are able to explain this behaviour through the off-axis injection of electron bunches linked to the oscillating asymmetry of the plasma wave with the CEP. However, the idealised case of a perfectly gaussian pulse fails to reproduce other observed properties, which are a smaller but existing variation of the beam pointing in the direction perpendicular to the polarization, and a significant charge evolution with the laser CEP. It is therefore necessary to take into account more complexity, such as experimental imperfections, to explain these features.\par

\subsection{Particle-in-cell simulations with asymmetric laser spot}
\label{sec:perpasym}

\begin{figure}[t]
    \includegraphics[width=0.48\textwidth]{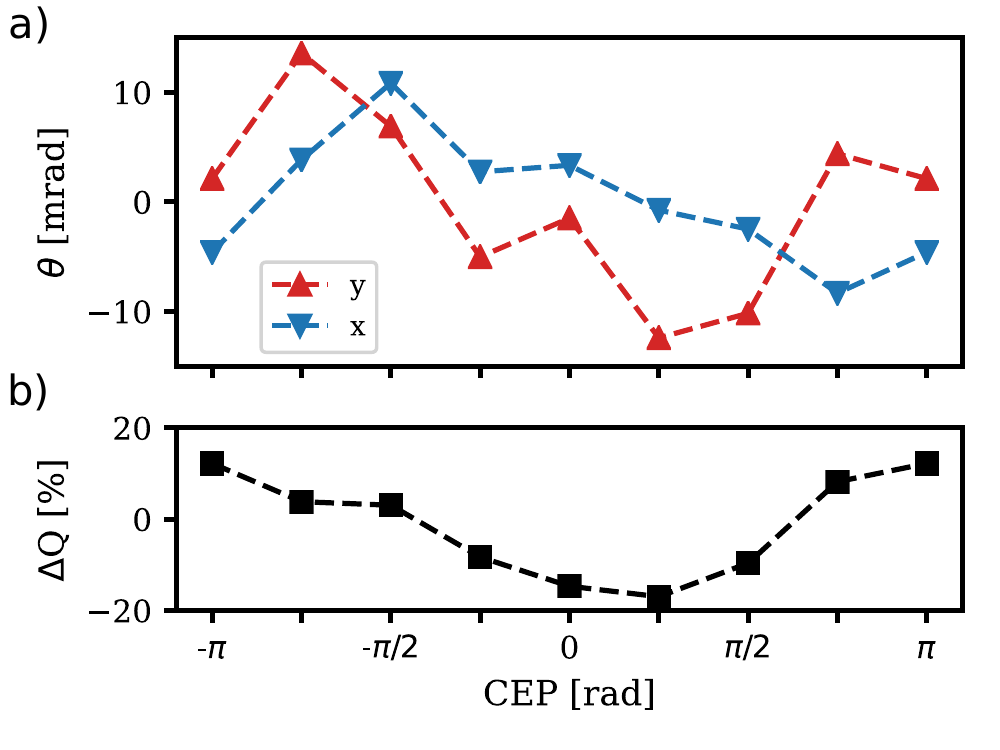}
    \caption{\label{fig:pointingasym} a) Beam pointing in the simulations for different CEPs, in the y (polarization) direction (red triangles), in the x direction (blue triangles).   b) Charge variation in the simulations plotted against the initial laser CEP. The mean charge $(\Delta Q = 0)$ corresponds to Q = 3.6\,pC.}
\end{figure} 

Beam pointing variations in the direction perpendicular to the laser polarization cannot be explained solely by an effect of the carrier-envelope phase. Indeed, the  plasma wave asymmetry due to the near-single cycle pulse develops only in the polarization direction and leaves the perpendicular plane perfectly symmetric. Another source of asymmetry, coupled with CEP must therefore be responsible for this behaviour. We have remarked in section\,\ref{sec:exp_res} that the laser focal spot corresponding to the experimental variations of $\theta_x$ with CEP is asymmetric in the horizontal direction, and therefore could be the source of this observation. To investigate this, we carried out PIC simulations in similar conditions as in Sec.\,\ref{sec:simsym}, but now using the measured experimental laser spatial profile. The laser energy is slightly re-adjusted to 2.9\,mJ. In these simulations, the accelerated charge is around 3.6\,pC with energies in the range 5-7\,MeV. \par
Firstly, these simulations reproduce the expected behaviour observed in the previous section with the beam pointing varying in the polarization direction when changing the initial laser CEP, as can be seen in Fig.\,\ref{fig:pointingasym}a. Additionally we observe that the pointing in the other direction is also evolving but with a smaller amplitude. This is interesting, not only because it reproduces the experimental observations, but because it indicates a coupling mechanism between the asymmetry induced by the laser spot in the perpendicular direction, and the one induced by the different CEPs in the polarization direction. Moreover, this coupling seems to have a significant effect on injection, because it induces large variations on the injected charge with total variations up to 30\% (see Fig.\,\ref{fig:pointingasym}b).\par

\begin{figure}[h!]
    \includegraphics[width=0.5\textwidth]{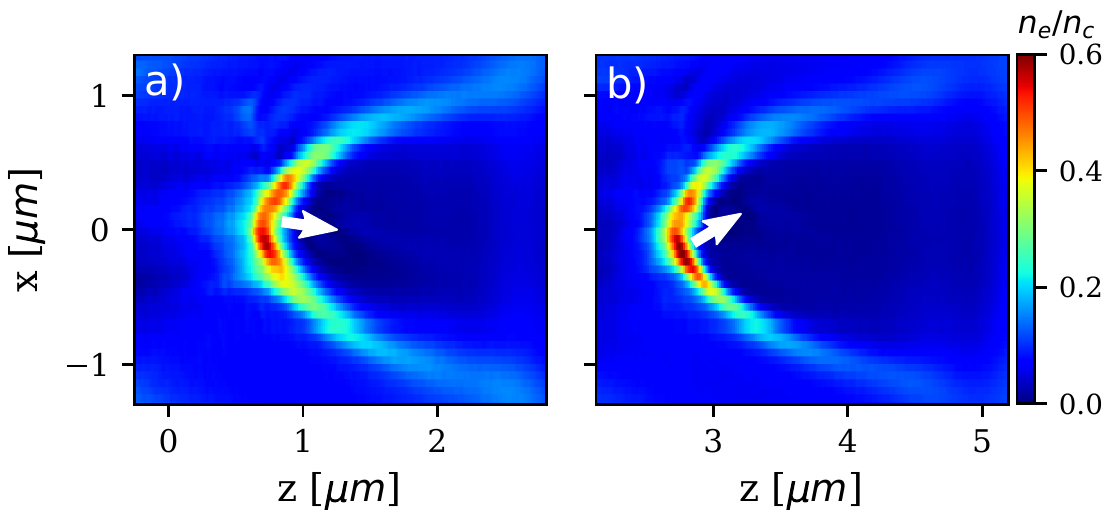}
    \caption{\label{fig:wakeperp} Slices of the electron density of the wake in the (x,z)-plane (perpendicular to polarization) for two injection times corresponding to two opposite CEP values. Injection occurs therefore at two different positions $y_1$ and $y_2$ due to CEP-induced off-axis injection    a) Slice for $y_1 = 0.2\mu m$ at a time where injection occurs from (y$>$0).  b)  Slice for $y_2 = -0.2\mu m$ at a time where injection occurs from (y$<$0). The two white arrows indicate the direction of injection of the electrons.}
\end{figure} 

In order to understand this coupling, we look at the geometry of the plasma wave and at the dynamics of the electrons in the perpendicular plane (x,z). Contrarily to the gaussian laser case, here the plasma wave also presents asymmetries in this plane, and these asymmetries lead to off-axis injection and residual transverse momentum $P_x$ for the injected electrons. Moreover, the bubble is a 3D structure, and these asymmetries vary depending on the positions in y at which the (x,z)-plane is observed. We have previously explained that the CEP-locked bubble oscillations lead to off-axis injection in the (y,z)-plane, on either positive or negative sides depending on the value of the CEP.
Therefore, when injected on different sides of the y-axis depending on the CEP, the electrons will see a different asymmetry in the (x,z)-plane due to the asymmetric laser spot and obtain a different transverse momentum $P_x$, explaining how the pointing in the perpendicular direction can be correlated to CEP. Figure\,\ref{fig:wakeperp} highlights this effect by showing two (x,z) slices of the plasma bubble for different times corresponding to two successive injection events associated with opposed CEPs. The slices are taken at position in y corresponding to the position of injection for each specific time. In the panel a) we observe a moderate upward asymmetry that will lead to injection of electrons with a low negative momentum, while in the panel b) the asymmetry is now leaning downwards and much more pronounced, and results in off-axis injection of electrons with a positive transverse momentum $P_x$. This complex coupling can also be responsible for variations in the injected charge, with situations where electrons are injected at a position in y where the spot-induced asymmetry leads to a locally higher amplitude of the wave and thus higher charge than at the opposed position at which electron would be injected with a different CEP.

\section{Conclusion}
These results demonstrate that the output beam parameters of a laser-plasma accelerator driven by a near-single cycle pulse are significantly impacted by the laser carrier-envelope phase. These effects can only be explained by going beyond the cycle-averaged ponderomotive approximation. It highlights the need to stabilize the CEP to achieve stable acceleration in both pointing and beam charge. Moreover, recent works propose that CEP-locked transverse bubble oscillation can play a role in LPA driven by initially longer but self-steepened pulses \cite{kim22_cep_arxiv,rakowski22_cep_arxiv}, indicating that the interest of CEP control might go beyond single-cycle pulses. Our findings also expand our previous work \cite{huijts22} by achieving a higher level of control through polarization, and confirm the physical explanation for the observed effects being attributed to a transverse oscillation of the wakefield, by obtaining them in helium which entirely exclude the ionization injection possibility.\par
Finally, our simulations indicate that the periodically injected bunches are particularly short, have low emittance and perform a collective betatron oscillation due to their off-axis injection. This makes the transverse bubble oscillation mechanism particularly interesting for enhanced ultrashort betatron radiation source \cite{kim21_cep,kim22_cep_arxiv,rakowski22_cep_arxiv} for X-ray production. 

\section*{Acknowledgements}
This work was funded by the Agence Nationale de la Recherche under Contract No. ANR-20-CE92-0043-01. Numerical simulations were performed using HPC resources from GENCI-TGCC (Grand Équipement National de Calcul Intensif) (Grant No. 2020-A0090510062) with the IRENE supercomputer. This project has also received funding from the European Union’s Horizon 2020 Research and Innovation program IFAST under Grant Agreement No. 101004730.

\section*{Competing Interests}
The authors declare no competing interests.

\section*{Data Availability}
The data that support the findings of this study are available from the corresponding author upon reasonable request.
\onecolumn

\begin{appendices}

\section{PIC simulation results for an initial CEP of $\pi/2$ }
\label{app:sim}

\begin{figure*}[h!]
\begin{center}
    \includegraphics[width=0.8\textwidth]{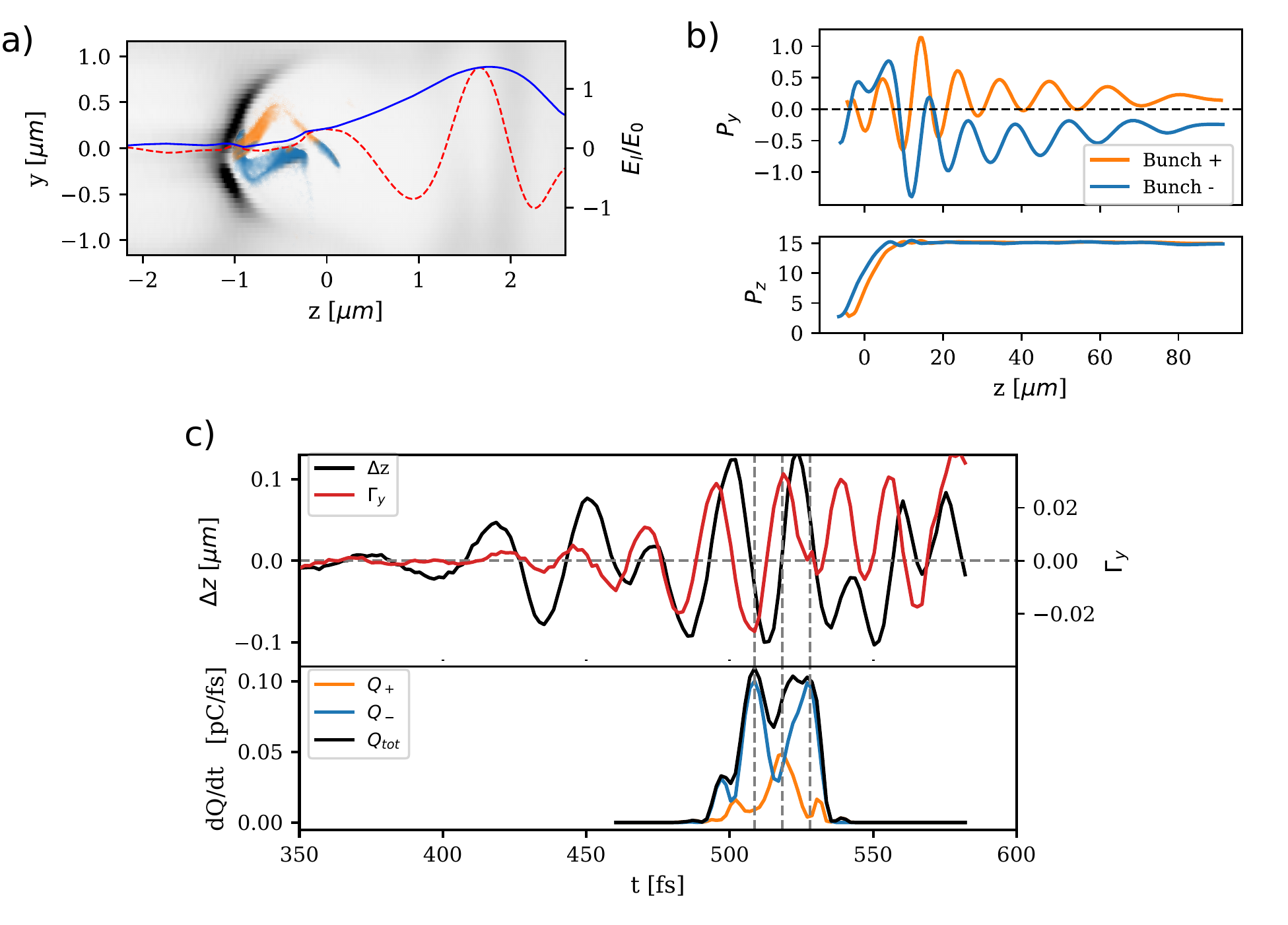}
    \caption{\label{figsup:simu_pi2} \textbf{Particle-in-cell simulation of a LPA driven by a 4.0\,fs laser in a helium plasma, for an initial CEP of $\pi/2$.}  a) Snapshot of the wakefield, showing three different injected sub-bunches. Electron density in the (y,z)-plane is shown in gray, and injected electrons are displayed in orange (blue) when their pointing is positive (negative) at the end of the simulation. The normalized laser electric field $E_l/E_0=E_l/(m_ec\omega_0/e)$ (red dashed line) and its envelope (blue solid line) are also shown.  b) Top : transverse momentum in the polarisation direction as a function of propagation distance for two sub-bunches injected with a positive (bunch +) and negative (bunch -) initial transverse momentum, corresponding to opposite signs of CEP-dependent bubble asymmetry at the moment of injection. Bottom: longitudinal momentum of the two sub-bunches. Both momenta are normalized to $\mathrm{m_ec}$.  c) Asymmetry of the wakefield in the y-direction (red) and difference in the mean position of the bubble in z between the top part (y$>$0) and the bottom part (y$<$0) of the wake (black). Charge injection rate for the two electron populations shown in a) with corresponding colors, as a function of the simulation time for an initial CEP of $\pi/2$. The gray dashed lines highlight the three main injection events.  }
\end{center}
\end{figure*}

Figure \ref{figsup:simu_pi2} shows a similar PIC simulation analysis as in Figure 6 of the main text, but with an initial CEP shifted by $\pi/2$. There are again clearly three main injection events, yielding three separate micro-bunches, but now electron are mainly injected from the top (blue) with a negative initial momentum and the total beam points downward at the end of the simulation.   

\end{appendices}

\newpage


\newpage

\begin{thebibliography}{10}
\expandafter\ifx\csname url\endcsname\relax
  \def\url#1{\burl{#1}}\fi
\expandafter\ifx\csname urlprefix\endcsname\relax\def\urlprefix{URL }\fi
\providecommand{\bibinfo}[2]{#2}
\providecommand{\eprint}[2][]{\url{#2}}
\providecommand{\doi}[1]{\url{https://doi.org/#1}}
\bibcommenthead

\bibitem{tajima_laser_1979}
\bibinfo{author}{Tajima, T.} \& \bibinfo{author}{Dawson, J.~M.}
\newblock \bibinfo{title}{Laser {Electron} {Accelerator}}.
\newblock \emph{\bibinfo{journal}{Phys. Rev. Lett.}}
  \textbf{\bibinfo{volume}{43}}~(4), \bibinfo{pages}{267--270}
  (\bibinfo{year}{1979}).
\newblock \urlprefix\url{https://link.aps.org/doi/10.1103/PhysRevLett.43.267}.
\newblock \doi{10.1103/PhysRevLett.43.267} .

\bibitem{gonsalves19}
\bibinfo{author}{Gonsalves, A.~J.} \emph{et~al.}
\newblock \bibinfo{title}{Petawatt laser guiding and electron beam acceleration
  to 8 gev in a laser-heated capillary discharge waveguide}.
\newblock \emph{\bibinfo{journal}{Phys. Rev. Lett.}}
  \textbf{\bibinfo{volume}{122}}, \bibinfo{pages}{084801}
  (\bibinfo{year}{2019}).
\newblock
  \urlprefix\url{https://link.aps.org/doi/10.1103/PhysRevLett.122.084801}.
\newblock \doi{10.1103/PhysRevLett.122.084801} .

\bibitem{rousse_production_2004}
\bibinfo{author}{Rousse, A.} \emph{et~al.}
\newblock \bibinfo{title}{Production of a {keV} {X}-{Ray} {Beam} from
  {Synchrotron} {Radiation} in {Relativistic} {Laser}-{Plasma} {Interaction}}.
\newblock \emph{\bibinfo{journal}{Phys. Rev. Lett.}}
  \textbf{\bibinfo{volume}{93}}~(13), \bibinfo{pages}{135005}
  (\bibinfo{year}{2004}).
\newblock
  \urlprefix\url{https://link.aps.org/doi/10.1103/PhysRevLett.93.135005}.
\newblock \doi{10.1103/PhysRevLett.93.135005}, \bibinfo{note}{publisher:
  American Physical Society} .

\bibitem{TaPhuoc2012}
\bibinfo{author}{Ta~Phuoc, K.} \emph{et~al.}
\newblock \bibinfo{title}{All-optical compton gamma-ray source}.
\newblock \emph{\bibinfo{journal}{Nature Photonics}}
  \textbf{\bibinfo{volume}{6}}~(5), \bibinfo{pages}{308--311}
  (\bibinfo{year}{2012}).
\newblock \urlprefix\url{https://doi.org/10.1038/nphoton.2012.82}.
\newblock \doi{10.1038/nphoton.2012.82} .

\bibitem{Chen2013}
\bibinfo{author}{Chen, S.} \emph{et~al.}
\newblock \bibinfo{title}{Mev-energy x rays from inverse compton scattering
  with laser-wakefield accelerated electrons}.
\newblock \emph{\bibinfo{journal}{Phys. Rev. Lett.}}
  \textbf{\bibinfo{volume}{110}}, \bibinfo{pages}{155003}
  (\bibinfo{year}{2013}).
\newblock
  \urlprefix\url{https://link.aps.org/doi/10.1103/PhysRevLett.110.155003}.
\newblock \doi{10.1103/PhysRevLett.110.155003} .

\bibitem{Wang2021}
\bibinfo{author}{Wang, W.} \emph{et~al.}
\newblock \bibinfo{title}{Free-electron lasing at 27 nanometres based on a
  laser wakefield accelerator}.
\newblock \emph{\bibinfo{journal}{Nature}}
  \textbf{\bibinfo{volume}{595}}~(7868), \bibinfo{pages}{516--520}
  (\bibinfo{year}{2021}).
\newblock \urlprefix\url{https://doi.org/10.1038/s41586-021-03678-x}.
\newblock \doi{10.1038/s41586-021-03678-x} .

\bibitem{mora_kinetic_1997}
\bibinfo{author}{Mora, P.} \& \bibinfo{author}{Antonsen, T.~M., Jr}.
\newblock \bibinfo{title}{Kinetic modeling of intense, short laser pulses
  propagating in tenuous plasmas}.
\newblock \emph{\bibinfo{journal}{Physics of Plasmas}}
  \textbf{\bibinfo{volume}{4}}~(1), \bibinfo{pages}{217--229}
  (\bibinfo{year}{1997}).
\newblock \urlprefix\url{https://doi.org/10.1063/1.872134} .

\bibitem{guenot_relativistic_2017}
\bibinfo{author}{Gu{\'e}not, D.} \emph{et~al.}
\newblock \bibinfo{title}{Relativistic electron beams driven by {kHz}
  single-cycle light pulses}.
\newblock \emph{\bibinfo{journal}{Nature Photon.}}
  \textbf{\bibinfo{volume}{11}}~(5), \bibinfo{pages}{293--296}
  (\bibinfo{year}{2017}).
\newblock \urlprefix\url{https://www.nature.com/articles/nphoton.2017.46}.
\newblock \doi{10.1038/nphoton.2017.46} .

\bibitem{rovige_demonstration_2020}
\bibinfo{author}{Rovige, L.} \emph{et~al.}
\newblock \bibinfo{title}{Demonstration of stable long-term operation of a
  kilohertz laser-plasma accelerator}.
\newblock \emph{\bibinfo{journal}{Phys. Rev. Accel. Beams}}
  \textbf{\bibinfo{volume}{23}}, \bibinfo{pages}{093401}
  (\bibinfo{year}{2020}).
\newblock
  \urlprefix\url{https://link.aps.org/doi/10.1103/PhysRevAccelBeams.23.093401}.
\newblock \doi{10.1103/PhysRevAccelBeams.23.093401} .

\bibitem{salehi2021laser}
\bibinfo{author}{Salehi, F.}, \bibinfo{author}{Le, M.},
  \bibinfo{author}{Railing, L.}, \bibinfo{author}{Kolesik, M.} \&
  \bibinfo{author}{Milchberg, H.}
\newblock \bibinfo{title}{Laser-accelerated, low-divergence 15-mev
  quasimonoenergetic electron bunches at 1 khz}.
\newblock \emph{\bibinfo{journal}{Physical Review X}}
  \textbf{\bibinfo{volume}{11}}~(2), \bibinfo{pages}{021055}
  (\bibinfo{year}{2021}).
\newblock \urlprefix\url{https://link.aps.org/doi/10.1103/PhysRevX.11.021055} .

\bibitem{he_electron_2012}
\bibinfo{author}{He, Z.-H.} \emph{et~al.}
\newblock \bibinfo{title}{Electron diffraction using ultrafast electron bunches
  from a laser-wakefield accelerator at khz repetition rate}.
\newblock \emph{\bibinfo{journal}{Applied Physics Letters}}
  \textbf{\bibinfo{volume}{102}}~(6), \bibinfo{pages}{064104}
  (\bibinfo{year}{2013}).
\newblock \urlprefix\url{https://doi.org/10.1063/1.4792057}.
\newblock \doi{10.1063/1.4792057} .

\bibitem{he_capturing_2016}
\bibinfo{author}{He, Z.-H.} \emph{et~al.}
\newblock \bibinfo{title}{Capturing structural dynamics in crystalline silicon
  using chirped electrons from a laser wakefield accelerator}.
\newblock \emph{\bibinfo{journal}{Scientific Reports}}
  \textbf{\bibinfo{volume}{6}}~(1), \bibinfo{pages}{36224}
  (\bibinfo{year}{2016}).
\newblock \urlprefix\url{https://doi.org/10.1038/srep36224}.
\newblock \doi{10.1038/srep36224} .

\bibitem{faure_concept_2016}
\bibinfo{author}{Faure, J.} \emph{et~al.}
\newblock \bibinfo{title}{Concept of a laser-plasma-based electron source for
  sub-10-fs electron diffraction}.
\newblock \emph{\bibinfo{journal}{Phys. Rev. Accel. Beams}}
  \textbf{\bibinfo{volume}{19}}~(2), \bibinfo{pages}{021302}
  (\bibinfo{year}{2016}).
\newblock
  \urlprefix\url{https://link.aps.org/doi/10.1103/PhysRevAccelBeams.19.021302}.
\newblock \doi{10.1103/PhysRevAccelBeams.19.021302} .

\bibitem{hidding_laser-plasma_2017}
\bibinfo{author}{Hidding, B.} \emph{et~al.}
\newblock \bibinfo{title}{Laser-plasma-based space radiation reproduction in
  the laboratory}.
\newblock \emph{\bibinfo{journal}{Scientific Reports}}
  \textbf{\bibinfo{volume}{7}}~(1), \bibinfo{pages}{42354}
  (\bibinfo{year}{2017}).
\newblock \urlprefix\url{https://doi.org/10.1038/srep42354}.
\newblock \doi{10.1038/srep42354} .

\bibitem{Rigaud2010}
\bibinfo{author}{Rigaud, O.} \emph{et~al.}
\newblock \bibinfo{title}{Exploring ultrashort high-energy electron-induced
  damage in human carcinoma cells}.
\newblock \emph{\bibinfo{journal}{Cell Death {\&} Disease}}
  \textbf{\bibinfo{volume}{1}}~(9), \bibinfo{pages}{e73--e73}
  (\bibinfo{year}{2010}).
\newblock \urlprefix\url{https://doi.org/10.1038/cddis.2010.46}.
\newblock \doi{10.1038/cddis.2010.46} .

\bibitem{lundh2012}
\bibinfo{author}{Lundh, O.} \emph{et~al.}
\newblock \bibinfo{title}{Comparison of measured with calculated dose
  distribution from a 120-mev electron beam from a laser-plasma accelerator}.
\newblock \emph{\bibinfo{journal}{Medical Physics}}
  \textbf{\bibinfo{volume}{39}}~(6Part1), \bibinfo{pages}{3501--3508}
  (\bibinfo{year}{2012}).
\newblock
  \urlprefix\url{https://aapm.onlinelibrary.wiley.com/doi/abs/10.1118/1.4719962}.
\newblock \doi{https://doi.org/10.1118/1.4719962},
  \bibinfo{eprint}{{\href{https://arxiv.org/abs/https://aapm.onlinelibrary.wiley.com/doi/pdf/10.1118/1.4719962}{{https://aapm.onlinelibrary.wiley.com/doi/pdf/10.1118/1.4719962}}}}
  .

\bibitem{cavallone2021}
\bibinfo{author}{Cavallone, M.} \emph{et~al.}
\newblock \bibinfo{title}{Dosimetric characterisation and application to
  radiation biology of a khz laser-driven electron beam}.
\newblock \emph{\bibinfo{journal}{Applied Physics B}}
  \textbf{\bibinfo{volume}{127}}~(4), \bibinfo{pages}{57}
  (\bibinfo{year}{2021}).
\newblock \urlprefix\url{https://doi.org/10.1007/s00340-021-07610-z}.
\newblock \doi{10.1007/s00340-021-07610-z} .

\bibitem{audet2021}
\bibinfo{author}{Audet, T.~L.} \emph{et~al.}
\newblock \bibinfo{title}{Ultrashort, mev-scale laser-plasma positron source
  for positron annihilation lifetime spectroscopy}.
\newblock \emph{\bibinfo{journal}{Phys. Rev. Accel. Beams}}
  \textbf{\bibinfo{volume}{24}}, \bibinfo{pages}{073402}
  (\bibinfo{year}{2021}).
\newblock
  \urlprefix\url{https://link.aps.org/doi/10.1103/PhysRevAccelBeams.24.073402}.
\newblock \doi{10.1103/PhysRevAccelBeams.24.073402} .

\bibitem{nerush_carrier-envelope_2009}
\bibinfo{author}{Nerush, E.~N.} \& \bibinfo{author}{Kostyukov, I.~Y.}
\newblock \bibinfo{title}{Carrier-{Envelope} {Phase} {Effects} in
  {Plasma}-{Based} {Electron} {Acceleration} with {Few}-{Cycle} {Laser}
  {Pulses}}.
\newblock \emph{\bibinfo{journal}{Phys. Rev. Lett.}}
  \textbf{\bibinfo{volume}{103}}~(3), \bibinfo{pages}{035001}
  (\bibinfo{year}{2009}).
\newblock
  \urlprefix\url{https://link.aps.org/doi/10.1103/PhysRevLett.103.035001}.
\newblock \doi{10.1103/PhysRevLett.103.035001} .

\bibitem{faure_review_2018}
\bibinfo{author}{Faure, J.} \emph{et~al.}
\newblock \bibinfo{title}{A review of recent progress on laser-plasma
  acceleration at {kHz} repetition rate}.
\newblock \emph{\bibinfo{journal}{Plasma Phys. Control. Fusion}}
  \textbf{\bibinfo{volume}{61}}~(1), \bibinfo{pages}{014012}
  (\bibinfo{year}{2018}).
\newblock \urlprefix\url{https://doi.org/10.1088%2F1361-6587%2Faae047}.
\newblock \doi{10.1088/1361-6587/aae047} .

\bibitem{huijts2021identifying}
\bibinfo{author}{Huijts, J.}, \bibinfo{author}{Andriyash, I.~A.},
  \bibinfo{author}{Rovige, L.}, \bibinfo{author}{Vernier, A.} \&
  \bibinfo{author}{Faure, J.}
\newblock \bibinfo{title}{Identifying observable carrier-envelope phase effects
  in laser wakefield acceleration with near-single-cycle pulses}.
\newblock \emph{\bibinfo{journal}{Physics of Plasmas}}
  \textbf{\bibinfo{volume}{28}}~(4), \bibinfo{pages}{043101}
  (\bibinfo{year}{2021}).
\newblock \urlprefix\url{https://doi.org/10.1063/5.0037925} .

\bibitem{xu_periodic_2020}
\bibinfo{author}{Xu, S.} \emph{et~al.}
\newblock \bibinfo{title}{Periodic self-injection of electrons in a few-cycle
  laser driven oscillating plasma wake}.
\newblock \emph{\bibinfo{journal}{AIP Advances}}
  \textbf{\bibinfo{volume}{10}}~(9), \bibinfo{pages}{095310}
  (\bibinfo{year}{2020}).
\newblock \doi{10.1063/5.0014691} .

\bibitem{kim21_cep}
\bibinfo{author}{Kim, J.}, \bibinfo{author}{Wang, T.}, \bibinfo{author}{Khudik,
  V.} \& \bibinfo{author}{Shvets, G.}
\newblock \bibinfo{title}{Subfemtosecond wakefield injector and accelerator
  based on an undulating plasma bubble controlled by a laser phase}.
\newblock \emph{\bibinfo{journal}{Phys. Rev. Lett.}}
  \textbf{\bibinfo{volume}{127}}, \bibinfo{pages}{164801}
  (\bibinfo{year}{2021}).
\newblock
  \urlprefix\url{https://link.aps.org/doi/10.1103/PhysRevLett.127.164801}.
\newblock \doi{10.1103/PhysRevLett.127.164801} .

\bibitem{huijts22}
\bibinfo{author}{Huijts, J.} \emph{et~al.}
\newblock \bibinfo{title}{Waveform control of relativistic electron dynamics in
  laser-plasma acceleration}.
\newblock \emph{\bibinfo{journal}{Phys. Rev. X}} \textbf{\bibinfo{volume}{12}},
  \bibinfo{pages}{011036} (\bibinfo{year}{2022}).
\newblock \urlprefix\url{https://link.aps.org/doi/10.1103/PhysRevX.12.011036}.
\newblock \doi{10.1103/PhysRevX.12.011036} .

\bibitem{lifschitz_optical_2012}
\bibinfo{author}{Lifschitz, A.~F.} \& \bibinfo{author}{Malka, V.}
\newblock \bibinfo{title}{Optical phase effects in electron wakefield
  acceleration using few-cycle laser pulses}.
\newblock \emph{\bibinfo{journal}{New J. Phys.}}
  \textbf{\bibinfo{volume}{14}}~(5), \bibinfo{pages}{053045}
  (\bibinfo{year}{2012}).
\newblock
  \urlprefix\url{https://doi.org/10.1088%2F1367-2630%2F14%2F5%2F053045}.
\newblock \doi{10.1088/1367-2630/14/5/053045} .

\bibitem{guenot_relativistic_2017-1}
\bibinfo{author}{Gu{\'e}not, D.} \emph{et~al.}
\newblock \bibinfo{title}{Relativistic electron beams driven by {kHz}
  single-cycle light pulses}.
\newblock \emph{\bibinfo{journal}{Nature Photon.}}
  \textbf{\bibinfo{volume}{11}}~(5), \bibinfo{pages}{293--296}
  (\bibinfo{year}{2017}).
\newblock \urlprefix\url{http://www.nature.com/articles/nphoton.2017.46}.
\newblock \doi{10.1038/nphoton.2017.46} .

\bibitem{bohle2014}
\bibinfo{author}{B{\"o}hle, F.} \emph{et~al.}
\newblock \bibinfo{title}{Compression of {CEP}-stable multi-{mJ} laser pulses
  down to 4{\hspace{0.167em}}fs in long hollow fibers}.
\newblock \emph{\bibinfo{journal}{Laser Phys. Lett.}}
  \textbf{\bibinfo{volume}{11}}~(9), \bibinfo{pages}{095401}
  (\bibinfo{year}{2014}).
\newblock \urlprefix\url{https://doi.org/10.1088/1612-2011/11/9/095401}.
\newblock \doi{10.1088/1612-2011/11/9/095401} .

\bibitem{ouille_relativistic-intensity_2020}
\bibinfo{author}{Ouill{\'e}, M.} \emph{et~al.}
\newblock \bibinfo{title}{Relativistic-intensity near-single-cycle light
  waveforms at {kHz} repetition rate}.
\newblock \emph{\bibinfo{journal}{Light. Sci. Appl.}}
  \textbf{\bibinfo{volume}{9}}~(1), \bibinfo{pages}{1--9}
  (\bibinfo{year}{2020}).
\newblock \urlprefix\url{https://www.nature.com/articles/s41377-020-0280-5}.
\newblock \doi{10.1038/s41377-020-0280-5} .

\bibitem{schmid2012supersonic}
\bibinfo{author}{Schmid, K.} \& \bibinfo{author}{Veisz, L.}
\newblock \bibinfo{title}{Supersonic gas jets for laser-plasma experiments}.
\newblock \emph{\bibinfo{journal}{Review of Scientific Instruments}}
  \textbf{\bibinfo{volume}{83}}~(5), \bibinfo{pages}{053304}
  (\bibinfo{year}{2012}).
\newblock \urlprefix\url{https://doi.org/10.1063/1.4719915} .

\bibitem{primot1995achromatic}
\bibinfo{author}{Primot, J.} \& \bibinfo{author}{Sogno, L.}
\newblock \bibinfo{title}{Achromatic three-wave (or more) lateral shearing
  interferometer}.
\newblock \emph{\bibinfo{journal}{JOSA A}} \textbf{\bibinfo{volume}{12}}~(12),
  \bibinfo{pages}{2679--2685} (\bibinfo{year}{1995}).
\newblock
  \urlprefix\url{http://opg.optica.org/josaa/abstract.cfm?URI=josaa-12-12-2679}
  .

\bibitem{primot2000extended}
\bibinfo{author}{Primot, J.} \& \bibinfo{author}{Gu{\'e}rineau, N.}
\newblock \bibinfo{title}{Extended hartmann test based on the pseudoguiding
  property of a hartmann mask completed by a phase chessboard}.
\newblock \emph{\bibinfo{journal}{Applied optics}}
  \textbf{\bibinfo{volume}{39}}~(31), \bibinfo{pages}{5715--5720}
  (\bibinfo{year}{2000}).
\newblock
  \urlprefix\url{http://opg.optica.org/ao/abstract.cfm?URI=ao-39-31-5715} .

\bibitem{kakehata2001single}
\bibinfo{author}{Kakehata, M.} \emph{et~al.}
\newblock \bibinfo{title}{Single-shot measurement of carrier-envelope phase
  changes by spectral interferometry}.
\newblock \emph{\bibinfo{journal}{Opt. Lett.}}
  \textbf{\bibinfo{volume}{26}}~(18), \bibinfo{pages}{1436--1438}
  (\bibinfo{year}{2001}).
\newblock
  \urlprefix\url{http://opg.optica.org/ol/abstract.cfm?URI=ol-26-18-1436} .

\bibitem{lucking2014approaching}
\bibinfo{author}{L{\"u}cking, F.}, \bibinfo{author}{Crozatier, V.},
  \bibinfo{author}{Forget, N.}, \bibinfo{author}{Assion, A.} \&
  \bibinfo{author}{Krausz, F.}
\newblock \bibinfo{title}{Approaching the limits of carrier-envelope phase
  stability in a millijoule-class amplifier}.
\newblock \emph{\bibinfo{journal}{Opt. Lett.}}
  \textbf{\bibinfo{volume}{39}}~(13), \bibinfo{pages}{3884--3887}
  (\bibinfo{year}{2014}).
\newblock
  \urlprefix\url{http://opg.optica.org/ol/abstract.cfm?URI=ol-39-13-3884} .

\bibitem{lehe_spectral_2016}
\bibinfo{author}{Lehe, R.}, \bibinfo{author}{Kirchen, M.},
  \bibinfo{author}{Andriyash, I.~A.}, \bibinfo{author}{Godfrey, B.~B.} \&
  \bibinfo{author}{Vay, J.-L.}
\newblock \bibinfo{title}{A spectral, quasi-cylindrical and dispersion-free
  {Particle}-{In}-{Cell} algorithm}.
\newblock \emph{\bibinfo{journal}{Comput. Phys. Commun.}}
  \textbf{\bibinfo{volume}{203}}, \bibinfo{pages}{66--82}
  (\bibinfo{year}{2016}).
\newblock
  \urlprefix\url{http://www.sciencedirect.com/science/article/pii/S0010465516300224}.
\newblock \doi{10.1016/j.cpc.2016.02.007} .

\bibitem{ammosov_tunnel_1986}
\bibinfo{author}{Ammosov, M.}, \bibinfo{author}{Delone, N.} \&
  \bibinfo{author}{Krainov, V.}
\newblock \bibinfo{title}{Tunnel ionization of complex atoms and of atomic ions
  in an alternating electromagnetic field}.
\newblock \emph{\bibinfo{journal}{Sov. Phys. JETP}}
  \textbf{\bibinfo{volume}{96}}~(6) (\bibinfo{year}{1986}).
\newblock \urlprefix\url{https://doi.org/10.1117/12.938695} .

\bibitem{bulanov_particle_1998}
\bibinfo{author}{Bulanov, S.}, \bibinfo{author}{Naumova, N.},
  \bibinfo{author}{Pegoraro, F.} \& \bibinfo{author}{Sakai, J.}
\newblock \bibinfo{title}{Particle injection into the wave acceleration phase
  due to nonlinear wake wave breaking}.
\newblock \emph{\bibinfo{journal}{Phys. Rev. E}}
  \textbf{\bibinfo{volume}{58}}~(5), \bibinfo{pages}{R5257--R5260}
  (\bibinfo{year}{1998}).
\newblock \urlprefix\url{https://link.aps.org/doi/10.1103/PhysRevE.58.R5257}.
\newblock \doi{10.1103/PhysRevE.58.R5257}, \bibinfo{note}{publisher: American
  Physical Society} .

\bibitem{tomassini_2003}
\bibinfo{author}{Tomassini, P.} \emph{et~al.}
\newblock \bibinfo{title}{Production of high-quality electron beams in
  numerical experiments of laser wakefield acceleration with longitudinal wave
  breaking}.
\newblock \emph{\bibinfo{journal}{Phys. Rev. ST Accel. Beams}}
  \textbf{\bibinfo{volume}{6}}, \bibinfo{pages}{121301} (\bibinfo{year}{2003}).
\newblock
  \urlprefix\url{https://link.aps.org/doi/10.1103/PhysRevSTAB.6.121301}.
\newblock \doi{10.1103/PhysRevSTAB.6.121301} .

\bibitem{beaurepaire_electron_2014}
\bibinfo{author}{Beaurepaire, B.}, \bibinfo{author}{Lifschitz, A.} \&
  \bibinfo{author}{Faure, J.}
\newblock \bibinfo{title}{Electron acceleration in sub-relativistic wakefields
  driven by few-cycle laser pulses}.
\newblock \emph{\bibinfo{journal}{New J. Phys.}}
  \textbf{\bibinfo{volume}{16}}~(2), \bibinfo{pages}{023023}
  (\bibinfo{year}{2014}).
\newblock
  \urlprefix\url{https://doi.org/10.1088%2F1367-2630%2F16%2F2%2F023023}.
\newblock \doi{10.1088/1367-2630/16/2/023023}, \bibinfo{note}{publisher: IOP
  Publishing} .

\bibitem{kim22_cep_arxiv}
\bibinfo{author}{Kim, J.}, \bibinfo{author}{Wang, T.}, \bibinfo{author}{Khudik,
  V.} \& \bibinfo{author}{Shvets, G.}
\newblock \bibinfo{title}{Polarization control of electron injection and
  acceleration in the plasma by a self steepening laser pulse}
  (\bibinfo{year}{2021}).
\newblock \urlprefix\url{https://arxiv.org/abs/2111.03014}.
\newblock \doi{10.48550/ARXIV.2111.03014} .

\bibitem{rakowski22_cep_arxiv}
\bibinfo{author}{Rakowski, R.} \emph{et~al.}
\newblock \bibinfo{title}{Transverse oscillating bubble enhanced laser-driven
  betatron x-ray radiation generation} (\bibinfo{year}{2022}).
\newblock \urlprefix\url{https://arxiv.org/abs/2202.01321}.

\end{thebibliography}
\end{document}